On a Generalized Mass-Velocity Relation for Disk Galaxies and Galaxy Clusters

Jeffrey M. La Fortune
1081 N. Lake St. Neenah, WI 54956 forch2@gmail.com
7 October 2021

*Abstract*
We consolidate the BFJ and BTF (MVD) relations into a generalized Scaling Mass Velocity Relation applicable to both pressure-supported galaxy clusters and rotation-supported galaxies. Unlike MOND inspired relations containing a characteristic acceleration scale and its normalization factor, our proposal is dependent on observed dynamic surface mass densities and discrepancies. We perform the same analysis for a sample of HIFLUGCS galaxy clusters (Tian et al, 2021) found in prior work with SPARC galaxies (arXiv: 2101.01537). For this galaxy cluster sample, we find little evidence for a universal acceleration constant as previously recognized for galaxies. We finish with an examination of the virial energies for the combined sample recovering a mass-energy relation consistent with the phenomenology.

*Introduction*
Galaxies and clusters exhibit phenomenal correlations (scaling relations) particularly for three fundamental dimensions, mass, length, and time. These are usually simple power laws, but describe specific self-similar properties that the structures tend to follow (Novak, 2012) (Zaritsky, 2008) (Schulz, 2017) (Stone, 2021) (Shenavar, 2021). Scaling relations may be presented as "tight" correlations with low residual scatter between parameters, while others may be significantly dispersed but still retaining general trends and central tendencies. Further, some scaling relations may also be transformed into analytical expressions when simple models are used, which is the approach taken here.

It seems with the introduction of dark matter and modified gravity a half-century ago, research directed toward classical dynamical models has been deemphasized. Somewhat disconcerting is that these mainstream models have made little progress *even* with the capacity to add mass or modify gravity to better match baryon dynamics. Here, we take a step back and model galaxy and clusters with four fundamental components; baryons, Newtonian dynamics, angular momentum all coupled via classical thermodynamics, specifically the virial theorem. Neither ΛCDM or MOdified Newtonian Dynamics (MOND) can offer a simpler explanation for the form and function of the mass-velocity relation proposed herein (Milgrom, 1983).

With recent advancements in precision astrometrics, data is now available for a comprehensive analysis for two prominent empirical mass-velocity correlations. First is the Baryonic Tully Fisher Relation (BTFR) specific to rotationally-supported disk galaxies and second is the Baryonic Faber Jackson Relation (BFJR) for velocity dispersion/pressure-dominated systems (Tully, 1977) (Faber, 1976). Our approach, termed the 'scaling model' has previously been used to analyze a large sample of SPARC galaxies and now, we apply the exact same model to a sample of 29 HIFLUGCS clusters (Tian, 2021 hereafter T21).

We review T21's modeling methodology used to estimate a Characteristic Acceleration Scale (CAS) for galaxy clusters based on MOND precepts. We interrogate HIFLUGCS galaxy cluster data using the proposed 'scaling model' and develop a BFTR zero-point expression that does not involve a pre-determined acceleration scale (La Fortune, 2020a) (La Fortune, 2021).

Next, we consolidate the BTF and BFJ into a single expression (gMVR) only using observed structural parameters. Lastly, we investigate the mass-energy relation for the combined SPARC and HIFLUGCS sample with results suggesting a link between the gMVR and equilibrium thermodynamics. Below, we recap Tian's analysis and comparison to similar studies.

*Galaxy Cluster MFJR / MVDR Relation – Tian 2021 (T21) Data and Model*
Recently, it has been demonstrated that a representative sample of 29 HIFLUGCS galaxy clusters follow the Mass Velocity Dispersion Relation, or MVDR. In this study, the MVDR is modeled in the "logarithmic plane" employing Bayesian/MCMC methodology neatly described in the publication with overall best fit:

$$\log M_{Bar} = 4.1^{+0.4}_{-0.4} \log \sigma_{last} + 1.6^{+1.0}_{-1.3} \qquad M_{Bar}\,[M_\odot]\ and\ \sigma_{last}\,[kms^{-2}]$$

Note that the intercept exhibits significant scatter ranging from 2 to 400 $M_\odot km^{-4}s^4$ and that the exponent is statistically equivalent to the MOND expectation value equal to four. Adding this constraint, the MVDR zero-point uncertainty is greatly reduced:

$$\log M_{Bar} = \mathbf{4} \log \sigma_l + 1.96^{+0.05}_{-0.06}$$

For this this analysis, the logarithmic form is converted to the equivalent power law expression for the remainder of this analysis:
$$M_{Bar} = 91.2 \sigma_l^4 \quad [M_\odot km^{-4} s^4][km^4/s^4]$$

The above equation is T21's best fit and is considered our 'exemplar' for comparison to the proposed 'scaling model.' Appendix A provides HIFLUGCS data and 'scaling' parameter statistics used in this analysis. Next, we briefly review the best fit zero-point normalization scheme used to extract a CAS estimate from the data. This is important as normalization is also employed in MOND to correct for galactic geometry in the form of a dimensionless constant (X=0.8) in order to obtain $a_0$=1.2x10$^{-13}$ kms$^{-2}$ (McGaugh, 2012).

*HIFLUGCS Cluster CAS Estimate and Comparison to Recent Studies*
T21's MONDian CAS estimate is dependent on the normalization factor employed in the MDVR zero-point expression. For this cluster sample, the authors opted for a form of normalization based on the Jeans factor ($J_\infty$) which is particularly appropriate for clusters modeled as isothermal spheres (Milgrom, 1984). The specific form used is Milgrom's $J_\infty=(\alpha_\infty-2\beta)^2$ with an asymptotic logarithmic density profile $\alpha_\infty$=4.86, and anisotropy coefficient β=0 (isotropic dispersion). With these values, the sample normalization factor is $J_\infty$ =23.7 with a cluster characteristic acceleration scale of g‡≈20x10$^{-13}$ kms$^{-2}$. Although this acceleration is ~20x $a_0$, it is consistent with Tian's previous 'RAR-inspired' analysis for a sample of 20 CLASH clusters (Tian, 2020). Below we compare the acceleration expressions between that of T21 and the galactic MOND mass-velocity relations:

$$T21\ HIFLUGCS\ Cluster\ CAS:\ g^\ddagger = \frac{J_\infty}{M_{Bar}G}\sigma_{last}^4; \quad MOND\ Galaxy\ CAS:\ a_0 = \frac{X}{M_{Bar}G}V_C^4$$

Note that implementation of $J_\infty$ for clusters is identical to MOND's normalization factor X for galaxies. Typical galaxy normalization is X≈0.8, but fails for clusters with spherical geometries having isotropic dispersion. For clusters, the MOND zero-point is predicted to be between $(a_0G)^{-1}$=63 and $(9/4)(a_0G)^{-1}$=141 based on 3D RMS space velocities (Milgrom, 1984). This prediction agrees with the HIFLUGCS best fit zero-point 91.2 $M_\odot km^{-4}s^4$.

MOND-based investigations have also attempted to 'test fit' the Radial Acceleration Relation (RAR) which is simply the correlation between baryonic and dynamic mass interior to $r_{last}$ (McGaugh, 2016). In an earlier investigation for a sample of 52 non-cooling core HIFLUGCS galaxy clusters, Chan obtained a best fit value $g^{\dagger}=9.3\pm1.7\times10^{-13}$ kms$^{-2}$ and found little correlation to the RAR (Chan, 2020).

More recently, an analysis of 12 Chandra and 12 X-COP galaxy clusters returned CAS values $g^{\dagger}=9.3\times10^{-13}$ kms$^{-2}$ and $11.2\times10^{-13}$ kms$^{-2}$ respectively with the RAR cleanly reproduced (Pradyumna, 2021a). Pradyumna also performed a follow up study with a selected 'Giles' sample of ten X-COP relaxed and non-cooled core clusters with $g^{\dagger}=15.9\pm1.4\times10^{-13}$ kms$^{-2}$ corroborating the previous estimate (Pradyumna, 2021b).

An informal census of current literature shows cluster CAS estimates are approximately an order of magnitude greater than MOND prescribed $a_0$. Additionally, these estimates vary greatly when compared to accelerations attributed to galaxies. In the next sections we investigate an alternative MVDR expression not dependent on MONDian or ΛCDM tenets, but relying only on the phenomenology, conventional physics, a simple and reasonable structural/kinematic model, and two structural parameters observed at $r_{last}$.

### 'Scaling Model' Cluster MVDR and HIFLUGCS Results

Using precision data commensurate with galaxies, we apply the same 'scaling' approach used in the SPARC analysis (La Fortune, 2021). The functional 'scaling' MVDR for the HIFLUGCS data set is shown below using T21 nomenclature. Terms are median 'central tendency' $CT_C$ [$M_\odot^{0.25}$/kms$^{-1}$], median dynamic surface density [$M_\odot$pc$^{-2}$] and mass discrepancy [$M_{\odot(Dyn)}/M_{\odot(Bar)}$]:

$$\text{Cluster Scaling "MVDR:"} \quad M_{Bar} = \frac{(CT_C)^4 \mu_{l(Med)}}{D\mu_{last}} \sigma_{last}^4$$

An important aspect of the 'scaling' zero-point is its dependency on two structural properties rather than a CAS/normalization factor. Plugging in the HIFLUGCS values, the 'scaling' mass-velocity relation retrieves the same zero-point value as T21's best fit:

$$\text{HIFLUGCS Scaling MVDR:} \quad M_{Bar} = \frac{4.60^4 \times 38.6}{D\mu_{last}} \sigma_{last}^4 = \frac{1.72 \times 10^4 M_\odot^2 s^4}{km^4 pc^2} \sigma_l^4 = 91.5\sigma_l^4$$

In this treatment, individual cluster zero-points are unique to each structure and are calculated directly from observed properties. In Appendix B, we derive the complementary galactic BFTR. Most serendipitously, we find that for their respective zero-points the numerators are numerically identical. This is the key that links the galactic and cluster mass-velocity relations under a single consolidated expression which is derived below.

### Consolidating the Mass-Velocity Relation for Galaxies and Clusters – the gMVR

The connection between the HIFLUGCS and SPARC data is found in the product of $CT^4$ and $\mu_{(Med)}$ which is equivalent to the dimensioned constant $1/\pi G^2$. This infers D and µ are independent of structure and a consolidated mass-velocity relation between structures. For reference, Appendix B summarizes SPARC 'scaling model' parameter values and associated statistics.

With a slight change in nomenclature, we arrive at the consolidated general Mass Velocity Relation or the 'scaling model 'gMVR:

$$\text{'Scaling' gMVR for Clusters and Galaxies:} \quad M_{Bar} = \frac{1}{\pi G^2 D \Sigma_{Dyn}} V^4$$

If a CAS is desired as is customary for MOND, it can be extracted by splitting the denominator into acceleration $\pi G D \Sigma_{Dyn}$ and G. Interestingly, this derivation had been reported over a decade ago for isothermal configurations (Sanders, 2010). Sanders communicated that for a constant surface density Σ and governed by the virial theorem, velocity dispersion scales as $M \approx (G^2 \Sigma)^{-1} \sigma^4$ and is identical to the empirically derived 'scaling' zero-point expression. Indeed, these so-called 'hidden variables,' or more appropriately zero-point terms, have been in plain sight all along (La Fortune, 2021).

Our gMVR two parameter solution parallels results obtained in a recent investigation linking dark matter halo properties with the galactic Radial Acceleration Relation (RAR) (Paranjape, 2021). In a mock set of nearly a quarter-million simulations, it was revealed that for a given baryonic mass, the scatter in the RAR can be entirely eliminated (made analytic) by incorporating two standard halo properties, virial mass ($m_{vir}$) and concentration ($c_{vir}$). These may be considered analogs to 'scaling' parameters mass discrepancy (D) and dynamic surface density ($\mu_{Obs}$), respectively.

In conventional RAR analyses where a constant CAS is invoked, the primary source of scatter is thought to be uncorrelated residuals (Lelli, 2016a). We counter this argument in that observed scatter in the gMVR represents *true* physical diversity in the population and can be treated as such. With the 'scaling' approach, we fully account for variables (see inset Figure 1) and construct an analytical model consistent with observed dispersion velocity.

Tian made an interesting point in his 2020 CLASH paper, positing that if baryons are coupled to equilibrium gas thermodynamics, then there should be a correlation between them and cluster dispersion. In fact, this correlation has been made explicit as the 'scaling' solution satisfies the virial theorem measured at outermost radii. In Appendix C, we calculate total virial energy as a function of baryon mass and obtain an unbroken power law consistent with the gMVR. Our results show that this coupling does not just apply to pressure-supported clusters but for rotationally-supported galaxies as well.

*HIFLUGCS Galaxy Cluster 'Scaling Model' Results*
In this section, we demonstrate the practical application of the scaling method in determining effective accelerations for each cluster. This investigation employs precisely the same model that mined SPARC galaxy data and figures in a recent pre-print (La Fortune 2021). We provide the same analysis, templates, and figures previously described to allow a transparent comparison between galaxy and cluster analysis. For the remaining sections, we follow the same discussion and figure format previously published for SPARC galaxies to invite visual 'side-by-side' analysis for the two classes of structure.

In Figure 1 we present the 'scaling' template for HIFLUGCS galaxy clusters plotting outermost radii ($r_{last}$) versus velocity dispersion ($\sigma_{last}$). It is at this radius the gMVR is specified. This figure is a three parameter deprojection with dynamic mass ($M_{Dyn}$) and surface mass density ($\mu_{last}$) constituting primary inputs for the 'scaling model.' The inset reproduces the HIFLUGCS MVDR identifying two clusters, Fornax and NGC 4636, illustrating their influence on the global fit.

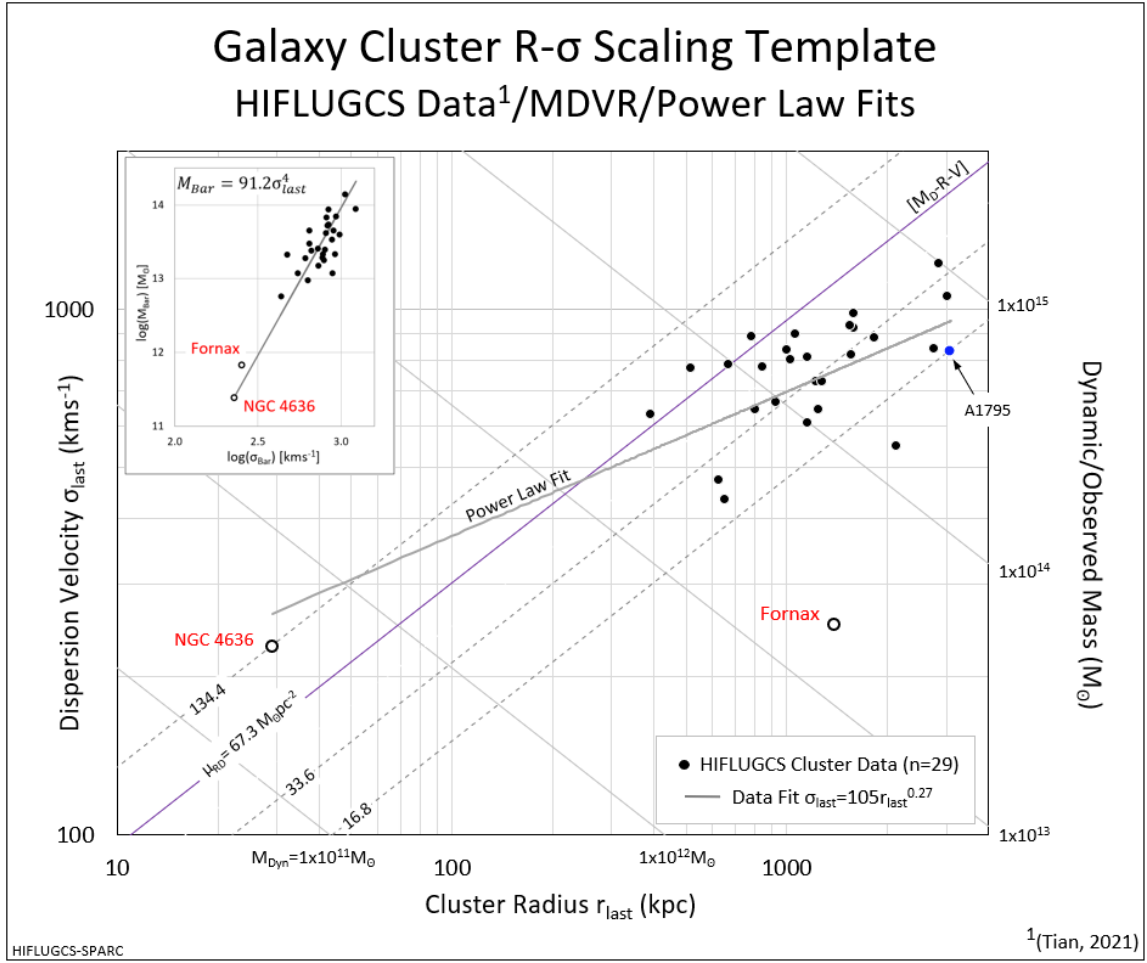

*Figure 1: HIFLUGCS cluster population (black points) on the r-σ plane with power law fit (light gray solid). The inset reproduces the best fit MVDR with dispersion exponent fixed to four. Fornax and NGC 4636 (open black circles) and cluster A1795 (blue solid) are identified for tracking purposes in later figures. The relatively shallow slope of the cluster fit is does not align with the [$M_D$-R-V] relation (purple solid) obtained from the SPARC data set. The surface density iso-lines (grey dashed) represent specific accelerations with the [$M_D$-R-V] relation equivalent to median $a_S=1.73\times10^{-13}$ $kms^{-2}$ found for the SPARC data. With surface density and edge acceleration linked as $M_D G r_l^{-2}=\sigma^2 r_l^{-1}$, the population of other RV product combinations 'off' this relation result in a Gaussian spread in the characteristic acceleration scale.*

Except for the Fornax and NGC 4636 clusters, the data is confined to a specific region in the of 'r-σ' plane. Fitting a power law results in relatively poor alignment with any surface mass (acceleration) isoline. Rather than the close alignment demonstrated by SPARC galaxies along the [$M_D$-R-V] relation that can be conceptually regarded as a CAS, HIFLUGCS clusters demonstrate no strong inclination. In the next figure, the same cluster data is presented in 'scaling model' D-μ parameter space. From a different perspective, we confirm very little cluster iso-acceleration alignment and weak evidence for a CAS.

In Figure 2 below, individual cluster mass discrepancies are plotted against dynamic surface mass densities – see Appendix A for modeling parameter definition and results. The inset demonstrates 1:1 'scaling' correspondence between observation and model (red points) compared with T21 best fit results (gray points).

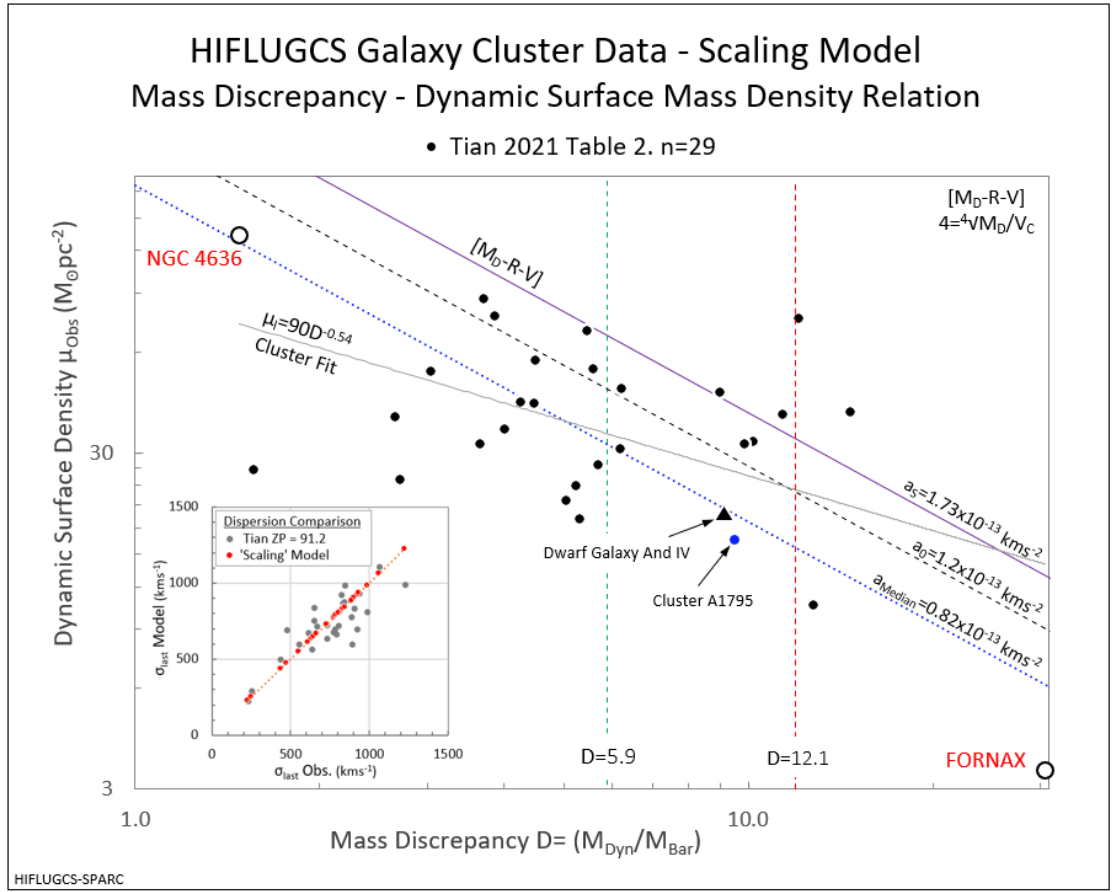

*Figure 2: HIFLUGCS clusters on the D-μ plane with power law fit (gray solid). Iso-accelerations are shown for Milgrom's $a_0$ (gray dash), galaxy [$M_D$-V-R] $a_S$ (purple solid), and cluster median $a_{Med}$ (blue dotted). NGC 4636 and Fornax flank opposing ends of the distribution and influences the power law slope away from the horizontal. The inset compares the individual dispersion fits for T21 (gray points) and 'scaling' (red points) demonstrating analytical nature of the model – see Appendix A. The proximity of dwarf galaxy Andromeda IV (black triangle) and cluster A1795 (blue point) represent similar global dynamics despite extreme differences in mass, morphology, and means of velocity support. – And IV source* (Karachentsev, 2016)

As in Figure 1, the cluster power law fit does not strongly align with any acceleration isoline. On the other hand, SPARC galaxy data strongly trends along [$M_D$-R-V] in the region *between* $a_0$ and $a_S$ for nearly the entire range of mass discrepancy. For galaxies, it is this alignment (with appropriate normalization) that delivers MOND CAS $a_0$=1.2x10$^{-13}$ kms$^{-2}$. While MOND may find utility with $a_0$ for rotationally-supported galaxies, we find the same treatment does not apply to galaxy clusters. For disk galaxies, we surmise that angular momentum is responsible for acceleration alignment that is absent in pressure-supported clusters.

For an interesting perspective, in Figure 2 we identify dwarf galaxy Andromeda IV sporting a very high gas fraction (>90%) similar to clusters. Despite substantial differences in mass and means of velocity support, And IV and A1795 may be paradoxically deemed dynamical 'twins' with much overlap between galaxies and clusters in 'D-μ' parameter space.

In Figure 3, we plot effective accelerations in ascending order following Rodrigues (Rodrigues, 2018) (Chang, 2019) (Marra, 2020). Note the change in acceleration units to ms$^{-2}$ following the original format. The SPARC galaxies are included with normalization factor X=0.7 for a median CAS equivalent to $a_0$.

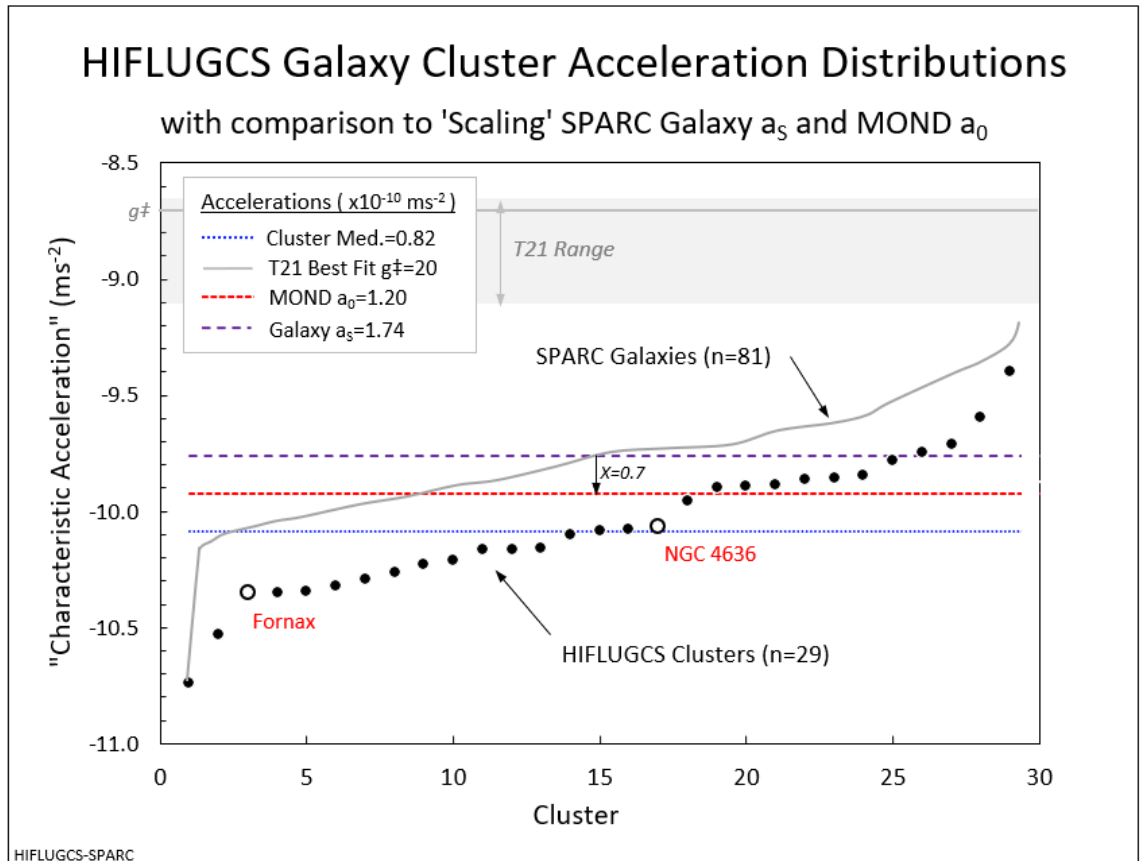

*Figure 3: HIFLUGCS cluster accelerations (black points) and median (blue dot) versus T21 best fit range $g‡=(8-22) \times 10^{-10}$ ms$^{-2}$ (gray shade) for normalization factor $J_\infty=23.7$. Note that T21's CAS $g‡$ is >10X MOND $a_0$ (red dash) and the SPARC galaxy median (purple dash). Except for the vertical offset, clusters follow the galaxy profile and remain bounded between similar min and max acceleration values. We illustrate the function of the MOND normalization factor X=0.7 that returns a CAS equal to Milgrom's constant $a_0$ for the SPARC galaxies.*

In the above figure, HIFLUGCS clusters demonstrate a similar acceleration distribution to that of SPARC galaxies, both in the trend and upper and lower bounds. The cluster median acceleration is approximately one-half the value of galactic effective acceleration $a_S$ and well below $a_0$. For clusters, any normalization factor must be greater than unity and to extract $a_0$ and find it to be X≈1.5, much lower than $J_\infty=23.7$ and attendant large $g‡$. Both Fornax and NGC 4636 deviate slightly from the smooth cluster acceleration trend, but are reasonably positioned within the range of 'scaling model' of the cluster acceleration distribution.

From a physical perspective, these low cluster accelerations may reflect prolate cluster geometries and/or highly anisotropic velocity dispersions, much different than any disk galaxy configuration. This 'scaling' derived normalization factor is consistent with Milgrom's MOND inspired estimate $1 \leq X \leq (9/4)$ for spherical isothermal systems discussed in a previous section.

*Generalized Scaling Model Relation/Cluster Data Comparison*

Below, we plot 'central tendency' ($CT_C$) for HIFLUGCS clusters as a Cumulative Distribution Function (CDF). This term, in combination with surface mass density links galaxies and clusters in the consolidated gMVR. We include the SPARC galaxy 'central tendency' ($CT_G$) distribution with its median 3.96 and standard deviation 0.58. Apart from the offset between $CT_G$ and $CT_C$, both distributions are highly Gaussian in nature. Note that the two low mass/dispersion galaxy clusters book-end the plot.

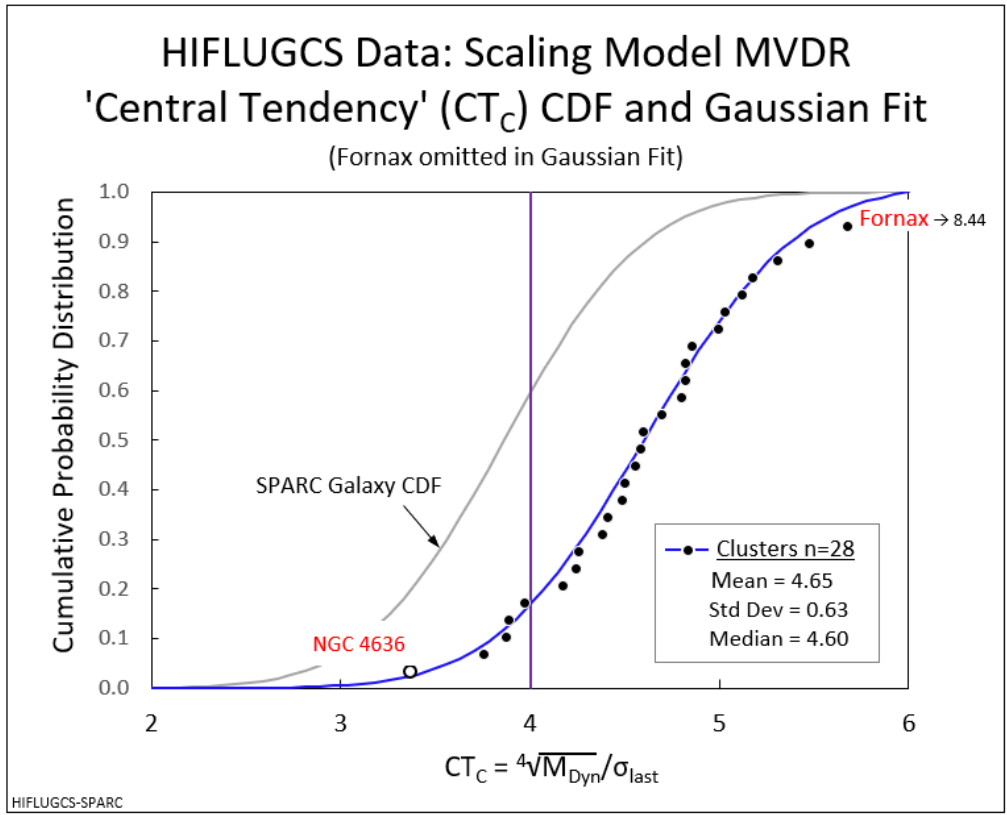

*Figure 4: HIFLUGCS cluster CDF versus the galactic 'scaling' central tendency value $CT_C \equiv 4$ (purple vertical). Omitting Fornax, the clusters present a Gaussian probability distribution (blue solid). The SPARC galaxy Gaussian CDF (gray solid) is included for comparison. For a given dynamic mass, clusters are offset to the right of galaxies indicating cluster dispersion is systematically lower than galactic rotation velocities. Fornax and NGC 4636 lie at opposing ends of the cluster data set with Fornax off the plot at $CT_C=8.44$.*

Above, we see the HIFLUGCS cluster 'central tendency ' data is well fit by a Gaussian CDF with little scatter. The similarities between cluster and galactic CDFs strongly suggest some form of global regulation as reflected by the mutual constant of proportionality $\pi G^2$ in the consolidated 'scaling' gMVR zero-point expression.

Note that Fornax is not included in the Gaussian fit. In Figure 1, this cluster lays significantly away from the data 'centroid.' Other than Fornax's unusual position in the "$r_l$-$\sigma_l$" plane and an excessively high $CT_C$, it remains consistent with the balance of the HIFLUGCS sample within the 'scaling' framework.

*Summary*

We offer a simple empirically-driven mass-velocity scaling relation characterized by two physical parameters; mass discrepancy D and dynamic mass surface density $\Sigma_{Dyn}$. For a fixed velocity exponent of four, we construct a generalized Mass-Velocity Relation (gMVR) consistent with first principles applicable to both disk galaxies and galaxy clusters. This model disputes the existence of natural acceleration constant $a_0$ finding its value physically dependent on the structural term $\pi GD\Sigma_{Dyn}$. We suggest the physical origin of the gMVR ($M \propto V^4$) is the virial theorem.

*Acknowledgements*

We would like to thank those who helped contributed to this manuscript and arXiv for open access to publish and have this work publicly available online.

*Appendix A: Table 1 - HIFLUGCS Cluster Data, Scaling Parameters, and Model Results*

| T21 HIFLUGCS Data (n=29)[1] | | | | | Scaling Model Parameters | | | | HIFLUGCS Galaxy Cluster Scaling Model Results | | | | | |
|---|---|---|---|---|---|---|---|---|---|---|---|---|---|---|
| Cluster Name | Baryon Mass | Velocity Dispersion | Terminal Radius | Baryon Mass | Baryon Dispersion | Baryon Accel. (point mass) | Dynamic Mass | "Simple" Zero Point Compare | Obs. Central Tendency | Dynamic SMD | Mass Discrepancy | Scaling gMVR Zero-Point ZP | Scaling Effective Accel. | Characteristic Accel. Scale (MOND X=1) |
| [ID] | log(M$_{Bar}$) [logM$_\odot$] | $\sigma_{last}$ [kms$^{-1}$] | $r_{last}$ [kpc] | M$_{Bar}$ [M$_\odot$] | $\sigma_{Bar}=\sqrt{(M_{Bar}G/r)}$ [kms$^{-1}$] | $a_{Bar}=\sigma_{Bar}^2/r$ [kms$^{-2}$] | M$_{Dyn}=R/\sigma^2/G$ [M$_\odot$] | M$_{Bar}/\sigma^4$ [M$_\odot$km$^{-4}$s$^4$] | $\sqrt[4]{(M_{Dyn})}/\sigma$ [M$_\odot^{0.25}$km$^{-1}$s] | $\mu=M_{Dyn}/\pi r^2$ [M$_\odot$pc$^{-2}$] | D=M$_{Dyn}$/M$_{Bar}$ | 4.60$^4$·38.6/D·$\mu$ [M$_\odot$km$^{-4}$s$^4$] | (ZP*G)$^{-1}$ [kms$^{-2}$] | "CAS" = X·(ZP·G)$^{-1}$ [kms$^{-2}$] |
| NGC 4636 | 11.379 | 229 | 29 | 2.39E+11 | 188 | 4.0E-14 | 3.54E+11 | 87.0 | 3.37 | 133.9 | 1.5 | 87.0 | 8.6E-14 | 8.6E-14 |
| Fornax | 11.827 | 252 | 1384 | 6.71E+11 | 46 | 4.9E-17 | 2.04E+13 | 166.5 | 8.44 | 3.4 | 30.4 | 166.5 | 4.5E-14 | 4.5E-14 |
| A3526 | 13.074 | 890 | 779 | 1.19E+13 | 256 | 2.7E-15 | 1.43E+14 | 18.9 | 3.89 | 75.3 | 12.1 | 18.9 | 4.0E-13 | 4.0E-13 |
| A1060 | 12.974 | 634 | 389 | 9.42E+12 | 323 | 8.7E-15 | 3.64E+13 | 58.3 | 3.87 | 76.5 | 3.9 | 58.3 | 1.3E-13 | 1.3E-13 |
| A262 | 13.070 | 551 | 2125 | 1.17E+13 | 154 | 3.6E-16 | 1.50E+14 | 127.5 | 6.35 | 10.6 | 12.8 | 127.5 | 5.9E-14 | 5.9E-14 |
| A3581 | 12.756 | 436 | 649 | 5.70E+12 | 194 | 1.9E-15 | 2.87E+13 | 157.8 | 5.31 | 21.7 | 5.0 | 157.8 | 4.8E-14 | 4.8E-14 |
| A4038 | 13.284 | 773 | 513 | 1.92E+13 | 401 | 7.13E-15 | 1.52E+14 | 53.9 | 3.76 | 86.2 | 3.7 | 53.9 | 1.4E-13 | 1.4E-13 |
| A2634 | 13.174 | 731 | 1220 | 1.49E+13 | 229 | 1.4E-15 | 1.52E+14 | 52.3 | 4.80 | 32.4 | 10.2 | 52.3 | 1.4E-13 | 1.4E-13 |
| A496 | 13.472 | 648 | 805 | 2.96E+13 | 398 | 6.4E-15 | 7.86E+13 | 168.2 | 4.60 | 38.6 | 2.7 | 168.2 | 4.5E-14 | 4.5E-14 |
| A2063 | 13.330 | 779 | 844 | 2.14E+13 | 330 | 4.2E-15 | 1.19E+14 | 58.1 | 4.24 | 53.2 | 5.6 | 58.1 | 1.3E-13 | 1.3E-13 |
| A2052 | 13.320 | 475 | 622 | 2.09E+13 | 380 | 7.5E-15 | 3.26E+13 | 410.4 | 5.03 | 26.9 | 1.6 | 410.4 | 1.8E-14 | 1.8E-14 |
| A2147 | 13.615 | 811 | 1146 | 4.12E+13 | 393 | 4.4E-15 | 1.75E+14 | 95.3 | 4.49 | 42.5 | 4.3 | 95.3 | 7.9E-14 | 7.9E-14 |
| A576 | 13.330 | 923 | 1582 | 2.14E+13 | 241 | 1.2E-15 | 3.13E+14 | 29.5 | 4.56 | 39.9 | 14.7 | 29.5 | 2.6E-13 | 2.6E-13 |
| A3571 | 13.734 | 841 | 999 | 5.42E+13 | 483 | 7.6E-15 | 1.64E+14 | 108.3 | 4.26 | 52.4 | 3.0 | 108.3 | 6.9E-14 | 6.9E-14 |
| A2589 | 13.279 | 610 | 1148 | 1.90E+13 | 267 | 2.0E-15 | 9.93E+13 | 137.3 | 5.18 | 24.0 | 5.2 | 137.3 | 5.5E-14 | 5.5E-14 |
| A2657 | 13.247 | 789 | 666 | 1.77E+13 | 338 | 5.5E-15 | 9.64E+13 | 45.6 | 3.97 | 69.2 | 5.5 | 45.6 | 1.6E-13 | 1.6E-13 |
| A119 | 13.652 | 648 | 1240 | 4.49E+13 | 394 | 4.1E-15 | 1.21E+14 | 254.5 | 5.12 | 25.1 | 2.7 | 254.5 | 3.0E-14 | 3.0E-14 |
| A3558 | 13.825 | 820 | 1558 | 6.68E+13 | 429 | 3.8E-15 | 2.44E+14 | 147.8 | 4.82 | 31.9 | 3.6 | 147.8 | 5.1E-14 | 5.1E-14 |
| A1644 | 13.649 | 901 | 1060 | 4.46E+13 | 425 | 5.5E-15 | 2.00E+14 | 67.6 | 4.17 | 56.7 | 4.5 | 67.6 | 1.1E-13 | 1.1E-13 |
| A3562 | 13.405 | 729 | 1269 | 2.54E+13 | 293 | 2.2E-15 | 1.57E+14 | 90.0 | 4.85 | 31.0 | 6.2 | 90.0 | 8.4E-14 | 8.4E-14 |
| A4059 | 13.378 | 666 | 926 | 2.39E+13 | 333 | 3.9E-15 | 9.55E+13 | 121.4 | 4.69 | 35.5 | 4.0 | 121.4 | 6.2E-14 | 6.2E-14 |
| A3391 | 13.525 | 885 | 1815 | 3.35E+13 | 282 | 1.4E-15 | 3.31E+14 | 54.6 | 4.82 | 31.9 | 9.9 | 54.6 | 1.4E-13 | 1.4E-13 |
| A85 | 13.843 | 934 | 1536 | 6.97E+13 | 442 | 4.1E-15 | 3.12E+14 | 91.5 | 4.50 | 42.0 | 4.5 | 91.5 | 8.2E-14 | 8.2E-14 |
| A133 | 13.392 | 803 | 1021 | 2.47E+13 | 322 | 3.3E-15 | 1.53E+14 | 59.3 | 4.38 | 46.8 | 6.2 | 59.3 | 1.3E-13 | 1.3E-13 |
| A3158 | 13.598 | 985 | 1581 | 3.96E+13 | 328 | 2.2E-15 | 3.57E+14 | 42.1 | 4.41 | 45.4 | 9.0 | 42.1 | 1.8E-13 | 1.8E-13 |
| A3266 | 13.942 | 1226 | 2848 | 8.75E+13 | 363 | 1.5E-15 | 9.96E+14 | 38.7 | 4.58 | 39.1 | 11.4 | 38.7 | 1.9E-13 | 1.9E-13 |
| A1795 | 13.716 | 831 | 3087 | 5.20E+13 | 269 | 7.6E-16 | 4.96E+14 | 109.0 | 5.68 | 16.6 | 9.5 | 109.0 | 6.9E-14 | 6.9E-14 |
| A2029 | 13.934 | 844 | 2751 | 8.59E+13 | 366 | 1.6E-15 | 4.56E+14 | 169.3 | 5.47 | 19.2 | 5.3 | 169.3 | 4.4E-14 | 4.4E-14 |
| A2142 | 14.142 | 1062 | 3012 | 1.39E+14 | 445 | 2.1E-15 | 7.90E+14 | 109.0 | 4.99 | 27.7 | 5.7 | 109.0 | 6.9E-14 | 6.9E-14 |
| Mean | | | | | | | | 107.9 | 4.78 | 42.6 | 7.05 | 107.9 | | |
| St. Dev. | | | | | | | | 79.1 | 0.94 | 26.4 | 5.69 | 79.1 | | |
| Median | | | | | | | | 91.5 | 4.60 | 38.6 | 5.31 | 91.5 | 8.2E-14 | 7.9E-14 |

[1] Tian et al, 2021 - Table 2. extract

*Appendix B: Scaling Model – Description and SPARC Galaxy Results (for Cluster comparison)*

The 'scaling model' framework is briefly discussed as an alternative to BTFR zero-point expressions relying on a characteristic acceleration scale (La Fortune, 2020a). The specifics are based on high-quality data from the SPARC disk galaxy data set (Lelli, 2016b) (La Fortune, 2021). The relation is shown on the left-hand side termed the 'central tendency' from which the general power-law 'scaling' relation is derived. The model base line with galactic dynamic surface mass density $\mu_{MRV}$=67.3 $M_\odot pc^{-2}$ and mass discrepancy D=5.9, returns a 'central tendency' precisely equal to four and is identified as the [$M_D$-R-V] relation in Figures 1, 2, and 4:

$$4 = CT_G = \frac{\sqrt[4]{M_{Dyn}}}{V_C} \quad or \quad M_{Dyn} = 256 V_C^4 \quad and \quad M_{Bar} = \frac{256}{D} V_C^4$$

Incorporating surface mass density, we arrive at the functional power-law expression shown below on the left. A comparison between the 'scaling' and MOND zero-point expressions show little commonality due to their differing foundational principles. The 'scaling' version is structurally dependent and varies with the physical properties of individual galaxies while MOND is confined to a single value for the entire population:

$$Galactic\ Scaling\ "BTFR" \quad M_{Bar} = \frac{(CT_G)^4 x \mu_{Median}}{D \mu_{RD}} V_C^4; \quad MOND\ BTFR \quad M_{Bar} = \frac{X}{a_0 G} V_{flat}^4$$

From SPARC data (as referenced in the main text), we recover a median central tendency $CT_G$=3.96 and an observed median dynamic surface mass density $\mu_{RD}$=69.8 $M_\odot pc^{-2}$. Plugging in these values, the galactic 'scaling' mass-velocity relation becomes:

$$SPARC\ Scaling\ "BTFR" \quad M_{Bar} = \frac{3.96^4 x 69.8}{D \mu_{RD}} V_C^4 = \frac{1.72 x 10^4\ M_\odot^2 s^4}{D \mu_{RD}} \frac{km^4}{km^4 pc^2} \frac{1}{s^4} = 43.3 V_C^4$$

The SPARC effective (observed) median acceleration is $1.74 \times 10^{-13}$ $kms^{-2}$. For a normalization factor X=0.7 we recover the universal acceleration scale $a_0$=$1.2 \times 10^{-13}$ $kms^{-2}$ – see Figure 3. Our median zero-point is in full agreement with McGaugh's result 47±6 $M_{Bar}km^{-4}s^4$ (for X=0.8) conducted some years earlier (McGaugh, 2012). Below, we summarize the galaxy statistics:

Table 2: Scaling Model Results for SPARC Galaxies (Q=1 cut, n=81) (La Fortune, 2021)

|  | "Simple" Zero Point Compare | Observed Central Tendency | Dynamic SMD | Mass Discrepancy | Scaling BTFR Zero-Point ZP | Scaling Effective Accel. | Characteristic Accel. Scale (X=0.7) |
|---|---|---|---|---|---|---|---|
|  | $M_{Bar}/V_C^4$ | $\sqrt[4]{(M_{Dyn})}/V_C$ | $\mu_{RD}=M_{Dyn}/\pi r R_D^2$ | $D=M_{Dyn}/M_{Bar}$ | $3.96^4 \cdot 69.8/D \cdot \mu_{RD}$ | $(ZP \cdot G)^{-1}$ | $CAS = a_{Dyn} \cdot X$ |
|  | [$M_\odot km^{-4}s^4$] | [$M_\odot^{0.25}km^{-1}s^2$] | [$M_\odot pc^{-2}$] | - - | [$M_\odot km^{-4}s^4$] | [$kms^{-2}$] | [$kms^{-2}$] |
| Mean | 53.2 | 3.90 | 99.7 | 5.42 | 53.6 | 1.40E-13 | 1.37E-13 |
| St. Dev. | 47.7 | 0.61 | 92.8 | 2.41 | 47.5 | 1.60E-13 | 8.21E-14 |
| Median | 42.8 | 3.96 | 69.8 | 5.01 | 43.3 | 1.74E-13 | 1.20E-13 |

The 'scaling model' assumes point mass baryon dynamics, constant circular velocity, and dynamic mass enclosed at outermost radii beyond which velocities decline in Keplerian fashion. This assumption is perhaps most debatable. Brouwer recently studied weak-lensing measurements beyond galactic disks to obtain velocity profiles into the extreme low acceleration regime roughly tracing the RAR (Brouwer, 2021). Our analysis of Milky Way dwarf galaxy satellites offers a counterpoint to this unsettled issue (La Fortune, 2020b).

*Appendix C: 'Scaling Model' Virial Theorem Relationships*

The 'scaling model' is founded on Newtonian gravity coupled with classical thermodynamics and satisfy the virial theorem. The theorem states for equilibrated structures, the time averaged potential energy is twice the kinetic energy:

$$\text{Kinetic Energy: } T = \frac{M_{Dyn}V^2}{2} \text{ and Potential Energy: } U = -\frac{GM_{Dyn}^2}{R}$$

$$\text{with: } 2T + U = 0; \; -T = \frac{1}{2U}; \; \text{or } U = 2E_{Tot}; \; E_{Tot} = -T$$

With algebra and substitution, the potential and kinetic energies can be recast with the gMVR zero-point expression in mind. The two energy components in T21 cluster nomenclature are:

$$U = -\frac{r_{last}\sigma_{last}^4}{G} = -\frac{r_l^2\sigma_l^2}{CT_C^4 G^2}\frac{\mu_{last}}{\mu_{Median}}; \quad T = \frac{r_{last}\sigma_{last}^4}{2G}$$

$$\text{Table 3 Equations:} \qquad\qquad (a) \qquad\qquad\qquad (b)$$

As an internal check, we ensure that the 'scaling' solutions are in agreement with the Peebles and Bullock spin equations (T = E$_{Tot}$) for spin parameter λ=0.423 (Peebles, 1971) (Bullock, 2001) (Marr, 2015):

$$\text{Equating } \lambda_P = \frac{J\sqrt{|E|}}{M_{Dyn}^{5/2}G} \text{ with } \lambda_B = \frac{J}{\sqrt{2}\sigma_{last}r_{last}} \rightarrow |E_{Tot}| = \frac{G^2 M_{Dyn}^3}{2\sigma_l^2 r_l^2} \quad (c)$$

In Table 3, selected global properties are listed for And IV and A1795 along with calculated virial energy content. The Milky Way is included as an internal check against previous energy estimates obtained from equations (a), (b), and (c) above.

Table 3: Properties and Energy Content for Cluster A1795 and Galaxies Andromeda IV and the Milky Way

| Structure Type | Structure Name | Terminal Velocity $\sigma_{last}$ & $V_C$ | Terminal Radius $r_{last}$ & $R_D$ | Baryon Mass $M_{Bar}$ | Dynamic Mass $M_{Dyn}$ | Mass Discrepany D=$M_{Dyn}/M_{Bar}$ | Dynamic SMD μ | Potential Energy U (a) | Total Energy T (b) | Spin Eq. Energy $|E_{Tot}|$ (c) |
|---|---|---|---|---|---|---|---|---|---|---|
| -- | [ID] | [kms$^{-1}$] | [kpc] | [M$_\odot$] | [M$_\odot$] | [ratio] | [M$_\odot$pc$^{-2}$] | [M$_\odot$km$^2$s$^{-2}$] | [M$_\odot$km$^2$s$^{-2}$] | [M$_\odot$km$^2$s$^{-2}$] |
| Cluster | A1795 | 831 | 3087 | 5.20E+13 | 4.96E+14 | 9.5 | 16.6 | -3.41E+20 | 1.71E+20 | 1.71E+20 |
| LSB Dwarf | And IV | 45 | 7.5 | 3.90E+08 | 3.50E+09 | 9.1 | 20.0 | -7.17E+12 | 3.58E+12 | 3.58E+12 |
| HSB Disk | Milky Way* | 210 | 48.6 | 8.50E+10 | 5.00E+11 | 5.9 | 99.0 | -2.20E+16 | 1.10E+16 | 1.10E+16 |

\* Scaling Toy Model (La Fortune 2019)

In Figure 5 below, we present the baryon mass-energy relation for disk galaxies and galaxy clusters. Currently, it is not known if pressure-supported elliptical and dwarf galaxies also follow this relation, but work with high density stellar disks reveal a robust correlation between gravitational binding energy and structure that complements the 'scaling' approach (Shi, 2021). For reference, we include an earlier estimate of the total energy of the Milky Way galaxy (Ninkovic, 1992a) (Ninkovic, 1992b).

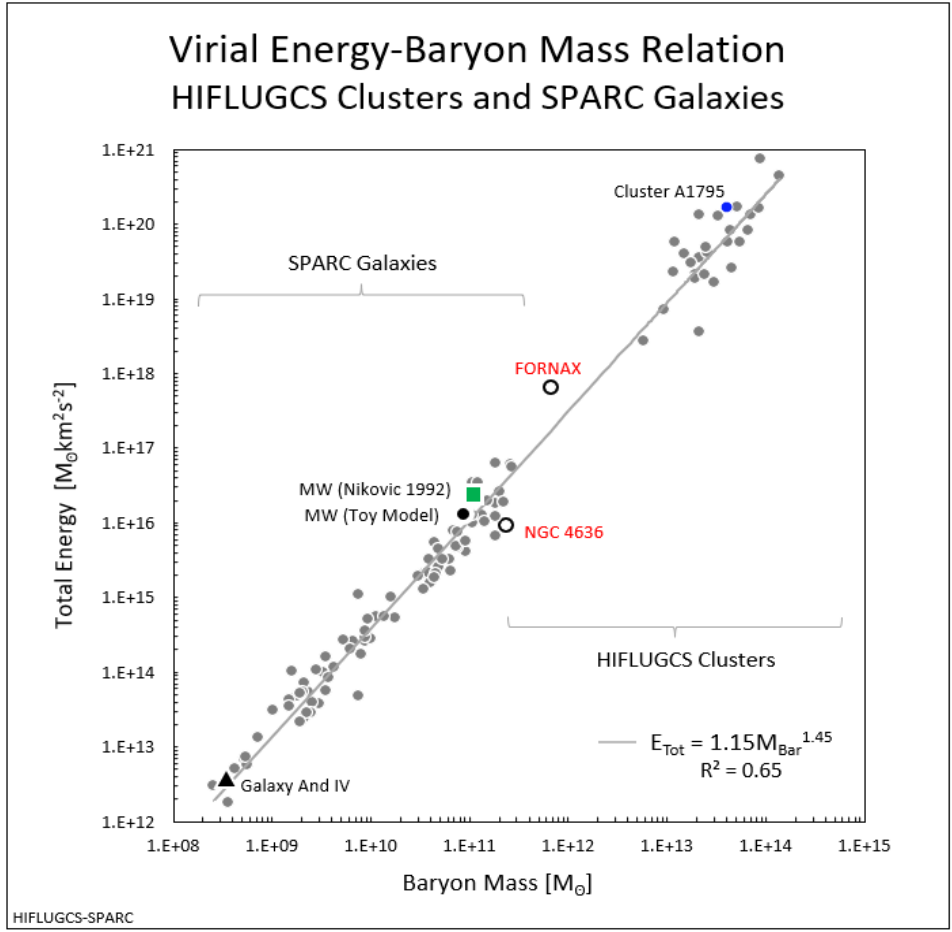

*Figure 5: The baryon mass - virial energy relation for combined HIFLUGCS and SPARC data (gray points) and power law fit $E_{Tot} \propto M_{Bar}^{3/2}$ (gray solid). Identified structures And IV (black triangle), A1795 (blue point) and two Milky Way models (green square and black point) are shown for perspective. T21 two low mass clusters, NGC4636 and Fornax bridge the relation where sparse data exists. Ninkovic's result for Milky Way total virial energy (green square) and toy 'scaling model' (black point) show close agreement.*

The Ninkovic model uses escape velocities as a means to constrain the Milky Way's gravitational potential and kinetic energy via the virial theorem and applied the concept of a corona to match Galaxy dynamics. Unlike more modern interpretations of a highly extended (>200 kpc) dark matter halo, Ninkovic's Galaxy corona is highly concentrated and shares similarities with the 'scaling' approach. As shown in Figure 2, galaxy cluster A1795 and dwarf galaxy Andromeda IV are included. Although they may be considered 'near-twins' in the D-μ plane, here they are at opposing ends of the spectrum. Conversely above, the Milky Way and NGC 4636 are close "mass-energy" neighbors despite their obvious differences.

We suggest the gMVR is a direct consequence of 'universal' physics in operation by equating $E \propto MV^2$ and $E \propto M^{3/2}$ with reduction to $M \propto V^4$ consistent with MOND. Most generally, this relation is at odds with the ΛCDM cosmologically motivated halo relation $M \propto V^3$ ($E_{Tot} \propto M^{5/3}$). We also find the inward extension of the gMVR essentially equivalent to the RAR (Wheeler, 2019). For constant circular velocity inside outermost radii, D and $\Sigma_{Dyn}$ are inversely proportional, contributing to the notable small scatter in this relationship (Lelli, 2016a).

Stone, C. (2021, May 3). The Intrinsic Scatter of Galaxy Scaling Relations. *ApJ, 12, 1*, pp. https://iopscience.iop.org/article/10.3847/1538-4357/abebe4/meta.

Tian, Y. (2020, Jan 23). The Radial Acceleration Relation in CLASH Galaxy Clusters. *arXiv*, p. https://arxiv.org/abs/2001.08340.

Tian, Y. (2021, Mar 25). Mass–Velocity Dispersion Relation in HIFLUGCS Galaxy Clusters. *ApJ, Vol. 1, 910, 56*, pp. https://iopscience.iop.org/article/10.3847/1538-4357/abe45c/meta.

Tully, R. B. (1977). A New Method of Determining the Distances to Galaxies. *A&A, 54, 661-673*, p. http://adsabs.harvard.edu/pdf/1977A&A....54..661T.

Wheeler, C. (2019, Sep 1). The Radial Acceleration Relation is a Natural Consequence of the Baryonic Tully-Fisher Relation. *ApJ, 842, 46*, pp. https://iopscience.iop.org/article/10.3847/1538-4357/ab311b/meta.

Zaritsky, D. (2008, Jul 20). Toward Equations of Galactic Structure. *ApJ, 682: 68-80*, p. http://iopscience.iop.org/article/10.1086/529577/meta.
16